\documentstyle[preprint,prl,aps,epsf]{revtex}
\begin{document}
\draft
\preprint{}
\title{Critical Properties of Random Quantum
Potts and Clock Models}
\author{T. Senthil and Satya N. Majumdar}
\address{Yale University\\
Department of Physics,  Sloane Physics Laboratory\\
New Haven, CT-06520-8120; USA.\\}
\date{20 October 1995}
\maketitle

\begin{abstract}
We study zero temperature phase transitions in two classes of random
quantum systems -the $q$-state quantum Potts and clock models.
For models with purely ferromagnetic interactions
in one dimension, we show that for strong randomness there
is a second order transition with critical properties that can
be determined exactly by use of an RG procedure. Somewhat surprisingly,
the critical behaviour is completely independent of $q$
(for $2 \leq q < \infty$).
For the $q > 4$ clock model, we suggest
the existence of a novel multicritical point
at intermediate randomness. We also consider the $T = 0$ transition
from a paramagnet to a spin glass in an infinite range model. Assuming
that the transition is second order, we solve for the critical behaviour
and find $q$ independent exponents.

\end{abstract}
\vspace{0.25cm}


\narrowtext

The effects of randomness on the properties of quantum many-body
systems have been the subject of experimental and
theoretical studies for many years\cite{MIT}. Despite this, understanding of
phenomena where quantum effects, randomness, and interactions are all
important is rather poor, and there are as yet few reliable
theoretical techniques. Recent work \cite{Fisher,Shankar,Read,Miller}
has focussed attention on
simple quantum statistical mechanical systems with randomness as a
useful starting point to obtain insights into such phenomena, and
some progress has been made.
In this paper we study analytically the effects of disorder on
the properties of two classes of quantum models with discrete symmetry.
These are to be regarded as
quantum versions of the classical $q$-state Potts
and the $q$-state clock models \cite{Wu}.
Studies of the corresponding classical models
have yielded a fair amount of insight into the competing
effects of randomness, interactions, and
thermal fluctuations\cite{Binder}. For instance,
classical Potts spin glasses have been studied
extensively as a paradigm for understanding the properties of orientational
glasses\cite{Binder}.

We first consider the zero temperature quantum phase transition
in random (purely ferromagnetic) $1$-dimensional quantum Potts and clock
chains.
For the $1$-d random transverse field
Ising model (RTFIM), which (as we shall see below) is the $q = 2$ case
of these models, a wealth of essentially exact information
has recently been obtained in a remarkable paper\cite{Fisher} by Fisher using
a real space renormalization group procedure. Here we show that this
procedure can similarly be used to obtain exact
critical properties of {\em strongly random}
$q$-state Potts and clock chains for all $q$, and implies,
remarkably, that there is no $q$ dependance in any of the exponents or the
scaling functions. This is in stark contrast to the
pure problem where the properties of the transition depend crucially
on the value of $q$\cite{Wu,Jose}. In addition, considerations on the
effects of weak randomness suggest the possibility
that for any amount of randomness, the Potts model for any $q$
and the clock model for $2 \leq q \leq 4$ are described by the
strong randomness fixed point. For
the clock chain with $q > 4$, we suggest the existence of a
multicritical point at a finite strength of randomness.
Next we consider the zero temperature transition from a
paramagnet to a spin glass in {\em infinite-ranged} quantum Potts and clock
models. Again building on work on the $q = 2$ case\cite{Miller,Read},
and assuming
a second order transition, we find that the critical properties
are independent of $q$.

The models are defined in terms of a variable that can assume
$q$ possible states (which we denote
$|0 \rangle~, |1 \rangle~,\cdots,|q - 1 \rangle$)
on the sites of a $d$-dimensional lattice.
The {\em classical} Potts (clock) interaction in the presence of a
uniform external ``magnetic'' field $H$ along the `$0$' direction is
\begin{eqnarray}
{\cal H}_{P,int} & = & - \sum_{\langle i, j \rangle} J_{ij}\delta_{n_i n_j}
                     - 2 H \sum_i(\delta_{n_i, 0} - \frac{1}{q})~~~~~~~~~~~~~
                     ~~~~~~~~~~(Potts) \\
{\cal H}_{C, int} & =  & - \sum_{\langle i,j \rangle} 2J_{ij}\cos(\frac{2
\pi}{q}
	    (n_i - n_j)) - 2 H \sum_i \cos\frac{2 \pi n_i}{q}~~~~~~~(Clock)
\end{eqnarray}
We introduce quantum fluctuations into these models by adding
at each site a ``transverse field'' term that attempts to change
the state of the variable at that site. Thus we consider
the quantum Hamiltonians
\begin{eqnarray}
\label{HamP}
{\cal H}_P & = & -\sum_i  h_i \left(\sum_{n_i,n'_i = 0}^{q - 1}\frac{1}{q}
                |n_i\rangle \langle n'_i|\right)
	            + {\cal H}_{P,int}
	    ~~~~~~~~~~~~~~~~~~~~~~~(Potts) \\
\label{HamC}
{\cal H}_C & = & -  \left(\sum_{n_i = 0}^{q - 1} h_i (|n_i \rangle \langle
n_i+1| +
          h.c )\right) + {\cal H}_{C,int}
	  ~~~~~~~~~~~~~~~~~~(Clock)
\end{eqnarray}
(We identify $ | n_i + q \rangle = | n_i \rangle $).
Through out the paper we will assume that the ${h_i}$ and ${J_{ij}}$
are independent random variables drawn from some distributions
$P_1(h)$ and $P_2(J)$ respectively. Note that at $H = 0$, the
Hamiltonian ${\cal H}_P$ is
invariant under a global
permutation $|n \rangle \rightarrow |n'\rangle$ of the states
at each site. For ${\cal H}_C$,
the symmetry is a global cyclic rotation $|n \rangle \rightarrow
|n + 1 \rangle$. Clearly for $q = 2$, both these models reduce to the
transverse
field Ising model. For general $q$, just as in the Ising case, the
``transverse field'' term plays the role of a kinetic energy that
opposes the tendency to order due to the interaction term. Also as in the
Ising case, the $d$-dimensional $q$-state quantum Potts (clock) model
Eqn.~\ref{HamP} (\ref{HamC})
at zero temperature
may be regarded as the transfer matrix in the
$\tau$-continuum limit of a $d + 1$-dimensional $q$-state
classical Potts (clock) model\cite{Pfeuty} with
disorder constant along one direction.

In the absence of disorder, the mapping to the classical $d + 1$-dimensional
pure problem provides a rather complete picture of the possible phases
and the transitions between them. For instance (at zero $H$),
the ferromagnetic ({\em i.e} $ J > 0,~~h > 0$)
quantum Potts chain has a first order transition for $q > 4$, and a
second order transition for $q \leq 4$ (for which all the exponents are
known exactly and depend on the value of $q$)\cite{Wu,den-Nijs}. The
ferromagnetic
clock chains, on the other hand, have, for $q > 4$, a quasi-long-range
ordered (QLRO) phase sandwiched between a truely long-range ordered phase
and a disordered phase\cite{Jose} . For $q \leq 4$, the quasi-long-range
ordered
phase disappears and is replaced by an ordinary second order phase transition
for which again all the exponents are known exactly\cite{Wu,Jose}. Below we
will
see that randomness drastically modifies this picture.

We start with $1$-d chains in which all the $h_i$'s and $J_{ij}$'s
are random but positive and $H = 0$. Defining the total magnetization
as $M = \sum_i \langle s_i \rangle$ with $s_i = \delta_{n_i,0} - \frac{1}{q}$
for the Potts model and $s_i = \cos\frac{2 \pi n_i}{q}$ for the clock model,
it is clear that as the overall relative strength of the $h_i$'s is decreased,
there will be a transition from a phase with $M = 0$ to one with $M \neq 0$.
We assume that the randomness is strong and follow
closely a real-space RG procedure used
by Fisher\cite{Fisher} to extract an enormous amount of information
on the RTFIM ($q = 2$ case). The basic idea behind this procedure\cite{Ma} is
to
successively eliminate the strongest coupling $\Omega = max\{h_i ,
J_{i, i + 1}\}$
in the chain  and get an effective Hamiltonian for the low energy degrees
of freedom.
First consider the case when the maximum coupling is a field, say $h_i$.
We eliminate the site $i$, and obtain, using second order perturbation
theory, a new effective bond between
the sites $i -1$ and $i + 1$ of strength
$J =  (J_{i-1,i} J_{i,i + 1})/ \kappa h_i$ where $\kappa$ is $q/2$ for
the Potts model and $(1 - \cos(\frac{2 \pi}{q}))/(1 + \delta_{q,2})$ for the
clock model. On the other hand, if the maximum coupling is a bond $J_{i, i +
1}$,
we replace the sites $i$ and $i + 1$ by a single Potts (or clock) degree of
freedom with an effective field $ h = (h_i h_{i + 1})/\kappa J_{i, i + 1}$
where $\kappa$ is the same as before (as may be expected from a duality
which these models can be shown to possess\cite{duality}). The $q$ dependance
is only
through $\kappa$. As in Ref.\cite{Fisher}, we convert these recursion
relations into flow equations for the distributions $P_1(\zeta = \ln(\Omega/J),
\Gamma = \ln( \Omega_I /\Omega))$  and $P_2 (\beta = \ln(\Omega/h),
\Gamma = \ln( \Omega_I/ \Omega))$ where $\Omega_I$ is the initial value of the
maximum coupling
and find
\begin{eqnarray}
\frac{\partial P_1}{\partial \Gamma}  = &&\frac{\partial P_1}{\partial \zeta}
 + P_1 ( P_1(0, \Gamma) - P_2(0, \Gamma)) + \nonumber \\
 &&~~~~~~+~~ P_2(0, \Gamma)\int d \zeta_1 d \zeta_2
 P_1(\zeta_1, \Gamma) P_1(\zeta_2, \Gamma)
\delta(\zeta - \zeta_1 -\zeta_2 - \ln\kappa)
\end{eqnarray}
and similarly with $P_1 \leftrightarrow P_2$ (as expected from duality).
When now we rescale and look for critical fixed points,
the value of $\ln \kappa$
becomes irrelevant at low energies (so long as it is finite). The resulting
probability distributions in the scaling limit described by the fixed
point are independent of $q$. In general it is necessary to keep track of
two joint distributions - that of bond lengths and bond strengths at scale
$\Gamma$ and that of cluster lengths, their magnetic moments and ``transverse''
field strengths at scale $\Gamma$. Both these distributions will be independent
of $q$ in the scaling limit.
{\em Thus the value of $q$ merely
determines a high-energy cutoff $E_c \sim \Omega_I e^{-|\ln\kappa|}$
(below which one should be
in order to observe scaling behaviour) but does not affect the
scaling behaviour itself.} Note that since this cutoff goes
to zero as $q \rightarrow \infty$, our
results hold only for finite $q$.

At this point it is tempting to conclude that all the Potts and clock chains
will have identical critical properties. However we note that, in general,
identical probability distributions do not necessarily imply identical
physical properties. For instance, the pure problems trivially have identical
distributions but have very different properties. However, as shown by
Fisher\cite{Fisher}, the distributions of the
logarithmic couplings $\zeta$ and $\beta$
become infinitely broad asymptotically at
low energies. It is then straightforward to see
that due to this extreme broadness all physical quantities (such as
magnetization or mean correlation function) are described by $q$-independent
scaling functions and exponents. The only $q$-dependence occurs in some
non-universal constants. We illustrate this point with the example
of the scaling of the magnetization $M(H, \delta)$ as a function of
small external applied field $H (> 0)$ and a dimensionless measure $\delta$
of the
deviation from criticality\cite{note}.
In the presence of a magnetic field $H$, the energy levels
of an otherwise-free cluster of magnetic moment $\mu$ split into a
ground state $|n = 0 \rangle$ and
other excited states with a gap $E_H = 2 \mu q H$ for the Potts case and
(atleast) $ E_H  = 2\mu H (1 - \cos \frac{2 \pi}{q})$ for the clock case.
Proceeding exactly as for the RTFIM, we stop the RG when the
maximum coupling $\Omega \sim E_H$. Due to the extreme broadness of
the distribution, an asymptotically exact expression for $M(H, \delta)$
is obtained by aligning all the remaining clusters in the direction of
the magnetic field. Thus
\begin{eqnarray*}
M(H, \delta) & = & \bar{\mu}\times({\rm total~~ number~~ of~~ active~~
({\em i.e}
              ~~undecimated)~~
               spins~~at~~scale~~} \Gamma_H = \ln\frac{D_H}{H})~~~ \\
             &   & +~~{\rm corrections}
\end{eqnarray*}
where $\bar{\mu}$ and $D_H$ are non-universal and possibly $q$ dependent
constants. The key point now is that the
number of active spins at a given scale $\Gamma$ is entirely a property
of the joint distribution of cluster lengths, magnetic moments, and field
strengths which is $q$-independent in the scaling limit. Consequently,
the universal scaling function describing
$M(H,\delta)$ will be the same for all $q$. The $q$ dependence is only
in the non-universal quantities $\bar{\mu}$ and $D_H$. Through similar
reasoning, one can establish that the mean correlation function
$\bar{C} ( x) = \overline{ \langle s_0 s_x \rangle}$  is also
described by a $q$-independent scaling function.

Thus the Potts and clock chains
for any $q$ (with strong ferromagnetic randomness) do indeed have critical
properties identical to those of the RTFIM (the $q = 2$ case)
for which detailed results are available\cite{Fisher}.
We point out some salient features below.
The spontaneous magnetization vanishes at the transition
with exponent $\beta = \frac{3 - \sqrt{5}}{2}$. The mean and
typical correlation functions
at the critial point decay as $\bar C(x) \sim \frac{1}{|x|^\beta}$
and $-\ln C_{typ}(x) \sim \sqrt|x|$ respectively for large $|x|$.
In the disordered side,
there are $2$ correlation lengths, characterizing the decay of mean and
typical correlations, which diverge with exponents $\nu = 2$ and
$\tilde \nu = 1$ respectively.
The magnetization scaling function is known exactly, and the scaling
function for the mean correlation function known up to
the solution of a linear ordinary differential equation.
Asymptotically at the critical point ``lengths'' scale as the square of the
logarithm of ``energies'' (unlike most other quantum transitions where lengths
scale as a power of the energies). On either side of the transition there
are Griffiths regions. In the Griffiths region of the disordered phase,
the order-parameter susceptibility diverges as $T \rightarrow 0$ as
a power with an exponent weaker than Curie.
Throughout this region, the magnetization increases as a power (with
logarithmic corrections)
of an applied external magnetic field with a continuously varying exponent.
Similarly in the Griffiths region of the ordered side
the stiffness vanishes for an infinite system, and the susceptibility
diverges as $T \rightarrow 0$ as a power with an exponent that is
stronger than Curie. Very far from the critical point, there are of course
the more conventional strongly ordered and disordered phases.

The RG procedure used above is valid only if the randomness in the
initial distributions $P_1$ and $P_2$ is strong. Clearly it cannot
address the question whether, if the initial distribution is narrow,
the low energy properties will still be described by the strong
randomness fixed point found above. Some insights on this matter
are provided by considering the effect of weak randomness on the
pure systems. So long as the pure transition is second order, weak
randomness is relevant at the fixed point if $\nu < 2/d$ (the generalized
Harris criterion\cite{Harris}) where $d$ is the spatial dimension.
{}From the known values
of $\nu$\cite{Wu,den-Nijs} for $q \leq 4$ for the Potts and clock chains,
we conclude that
weak randomness is indeed relevant. The simplest scenario then is that
for $q \leq 4$, the RG flows take the system to the strong randomness fixed
point for any amount of randomness in the initial distributions. The
situation however is different for $q > 4$. We discuss the Potts and clock
cases separately. The pure Potts chain for $q > 4$ has a first order
transition.
For first order phase transitions in classical systems, it has been argued
that any amount of ``bond'' randomness converts the transition to second
order\cite{Berker} for $d \leq 2 $. Extension of this argument to
quantum systems
would then suggest that the random $q > 4$ quantum Potts chain has a
second order transition for any amount of randomness. We conjecture that
this transition is described by the strong randomness fixed point.

\begin{figure}
\epsfxsize=8in
\centerline{\epsffile{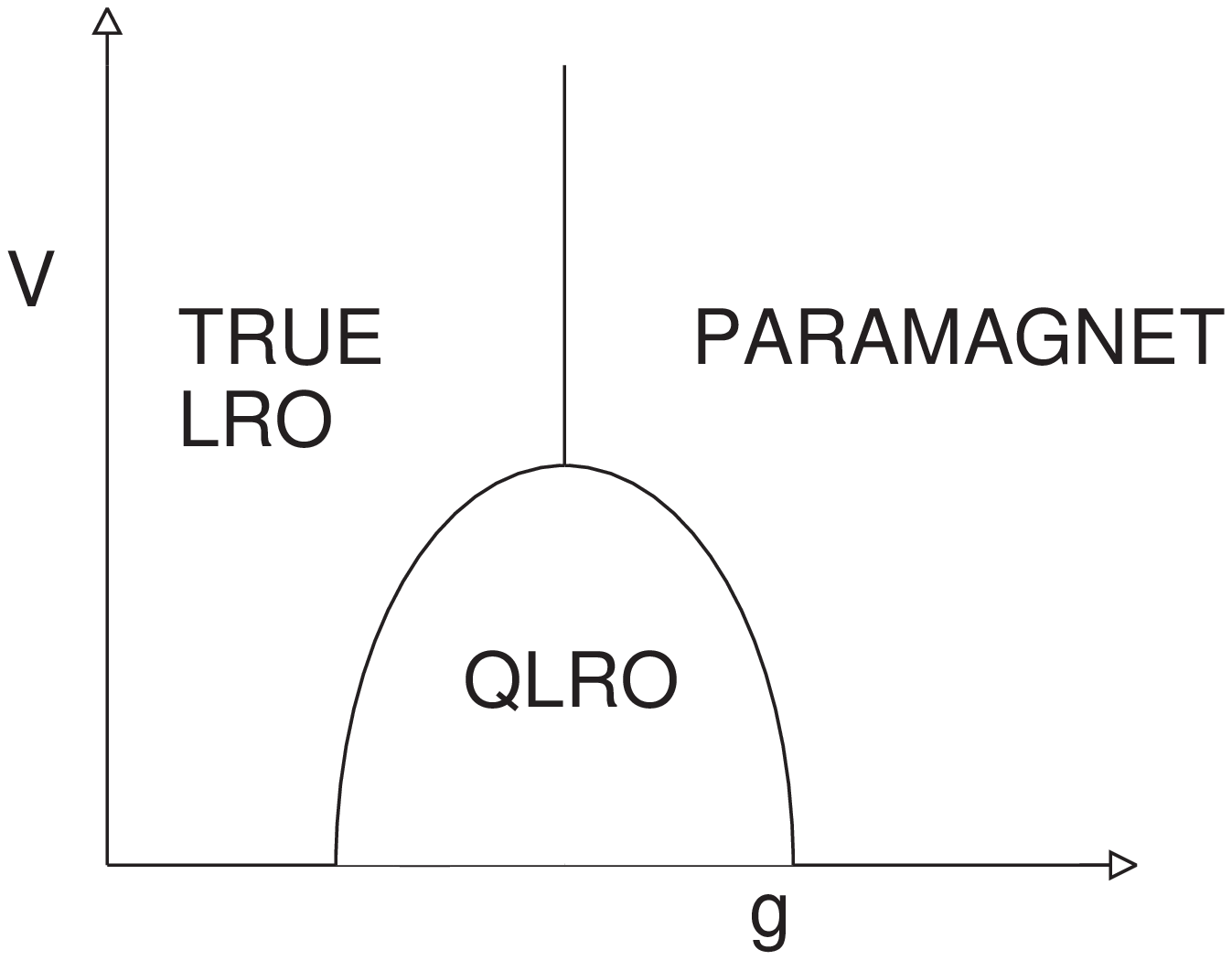}}
\vspace{0.5in}
\caption{One possible (schematic) phase diagram for the $q > 4$
random quantum clock chain. $g = \overline{\ln h}$ is a parameter
that measures the strength of quantum fluctuations and $V$ is a
measure of the strength of the randomness.}
\label{clock}
\end{figure}

We now turn to the clock model for $q > 4$ where there is an intermediate
quasi-long-range ordered phase separating a truely ordered phase from
the paramagnetic phase. The phase transitions into the QLRO phase from
either side are of the Kosterlitz-Thouless type with
$\nu = \infty$, therefore implying by the Harris criterion that weak randomness
is irrelevant. The QLRO phase is described by a line
of fixed points. A straightforward perturbative calculation shows that
weak randomness is irrelevant along this entire line\cite{Sen-Sat}.
Thus weak disorder
does not change the nature of the phase diagram in this case while for strong
disorder, as we have seen earlier, there is a single second order phase
transition (surrounded by Griffiths regions) and no QLRO phase.
Understanding how the phase diagram changes as the strength
of the disorder increases is an interesting open question. We speculate
that as the strength of the disorder is increased,
the two lines of Kosterlitz-Thouless transitions bounding the
QLRO phase merge at a multicritical point. Beyond this point the
QLRO phase disappears, and there is a single second order transition
described by the strong randomness fixed point (See Figure~\ref{clock}). We do
not
however have any strong arguments ruling out more complicated
scenarios (such as for instance, a new intermediate phase separating
the QLRO phase from the region with a single second order transition).

Having described the $1$-dimensional random ferromagnetic systems in some
detail, it is natural to ask if there are other non-trivial yet solvable cases.
One such example is provided by infinite-range spin glass models. As is
well-known, classical SG models display a complicated and rich structure
even for infinite-range interactions. There has been some recent
progress in understanding the zero temperature quantum phase transition
into a spin glass phase in the transverse field Ising model
(the $q = 2$ case of our models)\cite{Miller,Read}. The basic idea is quite
straightforward\cite{Bray}. One first performs the disorder average using the
replica trick and reduces the problem to an effective single site problem
with a self-consistency  condition on the auto-correlation function. At the
critical point it is possible to solve the self-consistency condition by
making a suitable ansatz for the long-time behaviour of the auto-correlation
function. For the Ising case, one finds that the critical auto-correlation
decays as $1/\tau^2$ at large imaginary time $\tau$\cite{Miller,Read}.
In addition, in the paramagnetic side, there is an energy gap that vanishes
on approaching the transition with an exponent $1/2$ with logarithmic
corrections.
Repeating the analysis for the $q$-state quantum Potts or clock spin
glass model, and assuming a second order transition, we find\cite{Sen-Sat}
that the critical
exponents once again remain the same as the Ising case. However the assumption
of a second order transition may be questionable (atleast
for the Potts and odd-$q$ clock models) since the finite
temperature phase transition from the paramagnetic side in the corresponding
classical models is not of the conventional second order type\cite{Gross}.

In summary, we have shown that for strongly random ferromagnetic quantum
$q$-state Potts and clock chains, the critical properties are
completely independent of $q$ for $2 \leq q < \infty$. This result is
{\em a priori} surprising as in the pure models the possible phases
and the transitions between them are known to depend crucially on the value
of $q$. We have also studied the zero temperature transition from a paramagnet
to a spin glass in these models
with infinite range interactions. If
the transition is second order, the critical exponents have $q$-independent
values. We conclude by noting the following implications of
our work and by raising some open questions.
Our results imply that
for classical $2$-d Potts and clock models with disorder correlated
along one direction,
the critical properties of the {\em finite temperature} phase transition
are independent of the value of $q$. It is interesting that numerical
simulations by Chen et. al.\cite{Landau} on the
classical $2$-d $8$-state Potts model with uncorrelated disorder find
a second order transition with exponents equal to the classical $2$-d Ising
values. This suggests the possibility that the $q$-independence found here
with correlated disorder also holds with uncorrelated disorder. However
an analytic calculation by expanding in $q - 2$ \cite{Ludwig} does find
$q$-dependence in the exponents. Further studies to address this issue
will be welcome.
{}From our discussion of the phase diagram for
the clock chains
with $q > 4$ (Figure~\ref{clock}), there arises the possibility of a novel
multicritical point at finite randomness. Verification of the existence of
this point, perhaps numerically, is an interesting open problem. Also
interesting is the question of whether this $q$ independence persists
in higher dimensional ferromagnetic models. For the spin glass models,
one approach to go beyond our results, atleast in high enough
dimensions, would be to study a
Landau theory analogous to that developed for the Ising and rotor
models\cite{Read}. Such a study may also resolve the issue of whether the
transition is second order or not.

We thank Subir Sachdev, R.Shankar, N.Read, and D.S.Fisher for useful
discussions
and comments. This research was supported by NSF Grants No. \ DMR-92-24290
and DMR-91-20525.

\begin{references}
\bibitem{MIT} An example is the metal-insulator transition.
For a review,
see P.A.Lee and T.V.Ramakrishnan, Rev. Mod. Phys. {\bf 57}, 287 (1985).
As another
example see work on the dirty boson problem by M.P.A.Fisher ,
P.Weichmann,
G.Grinstein, and D.S.Fisher, Phys. Rev. {\bf B40}, 546 (1989)

\bibitem{Fisher} D.S.Fisher,Phys. Rev. Lett. {\bf 69}, 534 (1992);
D.S.Fisher,Phys. Rev. {\bf B51}, 6411 (1995).

\bibitem{Shankar} R.Shankar and G.Murthy, Phys. Rev. B36, 536 (1987).

\bibitem{Read} N.Read, S.Sachdev, and J.Ye, Phys. Rev. {\bf B52}, 384 (1995);
J.Ye, S.Sachdev, and N.Read, Phys. Rev. Lett. {\bf 70}, 4011 (1993)

\bibitem{Miller} J.Miller and D.Huse, Phys. Rev. Lett. {\bf 70}, 3147 (1993)

\bibitem{Wu} F.Y.Wu, Rev. Mod. Phys. {\bf 54}, 235 (1982)

\bibitem{Binder}K.Binder and J.D.Reeger, Adv. in Physics, {\bf 41},547 (1992)

\bibitem{Jose}J.Jose, L.Kadanoff, S.Kirkpatrick, and D.R.Nelson, Phys.
Rev. {\bf B16}, 1217 (1977) ; S.Elitzur, R. Pearson, and J. Shigemitsu,
Phys. Rev. {\bf D19}, 3698 (1979)

\bibitem{Pfeuty} J.Solyom and P.Pfeuty, Phys. Rev. {\bf B 24}, 218 (1981);
L.Turban, J. Phys. A {\bf 18}, 2313 (1985)

\bibitem{den-Nijs}M.P.M. den-Nijs, J. Phys. {\bf A12}, 1857 (1979).

\bibitem{Ma}S.K.Ma, C.Dasgupta, and C.-k.Hu, Phys.Rev. Lett. {\bf 43},
1434 (1979); C.Dasgupta and S.K.Ma, Phys. rev. {\bf B22}, 1305 (1980)

\bibitem{duality}This is well-known for the pure $1$-d quantum systems;
see Ref~\cite{Wu}. In the presence of randomness, under the duality
transformation, the distributions of $h$ and $J$ get interchanged.

\bibitem{note}In Ref\cite{Fisher} for the RTFIM, $\delta$ was related exactly
to some properties of the original distributions. Such a relation does'nt
seem possible here.

\bibitem{Harris} A.B.Harris, J. Phys. C {\bf 7}, 1671 (1974); J.T.Chayes,
L.Chayes, D.S.Fisher, and T.Spencer, Phys. Rev. Lett. {\bf 57}, 2999 (1986).

\bibitem{Berker}K.Hui and A.N.Berker, Phys. Rev. Lett. {\bf 62}, 2507 (1989)

\bibitem{Sen-Sat}T. Senthil and S.N. Majumdar, unpublished

\bibitem{Bray}A.J.Bray and M.A.Moore, J. Phys. C {\bf 13}, L655 (1980)

\bibitem{Gross}D.J.Gross, I.Kanter, and H.Sompolinsky, Phys. Rev. Lett. {\bf
55},
304 (1985)


\bibitem{Landau} S.Chen, A.M.Ferrenberg, and D.P.Landau, Phys. Rev. Lett.,
{\bf 69}, 1213 (1992)

\bibitem{Ludwig} A.W.W.Ludwig, Nucl. Phys. {\bf B330}, 639 (1990)


\end{references}
\end{document}